\documentclass[twocolumn]{aastex631}
\usepackage{natbib}
\usepackage{booktabs}
\usepackage{multirow}

\newcommand {\msun} {M$_{\odot}$}
\newcommand{\degree} {$^\circ$}

\shorttitle{SMC X-1 in Stray Light}
\shortauthors{Brumback et al.}

\begin{document}

\title{Extending the baseline for SMC X-1's spin and orbital behavior with NuSTAR stray light}

\correspondingauthor{McKinley C.\ Brumback}
\email{mckinley@caltech.edu}

\author[0000-0002-4024-6967]{McKinley C.\ Brumback}
\affiliation{Cahill Center for Astronomy and Astrophysics, California Institute of Technology, 1216 E California Blvd, Pasadena, CA 91125, USA}

\author[0000-0002-1984-2932]{Brian W.\ Grefenstette}
\affiliation{Cahill Center for Astronomy and Astrophysics, California Institute of Technology, 1216 E California Blvd, Pasadena, CA 91125, USA}

\author[0000-0002-5341-6929]{Douglas J.\ K.\ Buisson}
\affiliation{Department of Physics and Astronomy, University of
Southampton, Highfield, Southampton, SO17 1BJ}

\author[0000-0002-4576-9337]{Matteo Bachetti}
\affiliation{INAF-Osservatorio Astronomico di Cagliari, via della Scienza 5, I-09047 Selargius (CA), Italy}

\author[0000-0002-8908-759X]{Riley Connors}
\affiliation{Cahill Center for Astronomy and Astrophysics, California Institute of Technology, 1216 E California Blvd, Pasadena, CA 91125, USA}

\author[0000-0003-3828-2448]{Javier A. Garc\'ia}
\affiliation{Cahill Center for Astronomy and Astrophysics, California Institute of Technology, 1216 E California Blvd, Pasadena, CA 91125, USA}

\author[0000-0002-3850-6651]{Amruta Jaodand}
\affiliation{Cahill Center for Astronomy and Astrophysics, California Institute of Technology, 1216 E California Blvd, Pasadena, CA 91125, USA}

\author[0000-0003-2737-5673]{Roman Krivonos}
\affiliation{Space Research Institute of the Russian Academy of Sciences (IKI), Moscow, Russia, 117997}

\author[0000-0002-8961-939X]{Renee Ludlam}
\altaffiliation{NASA Einstein Fellow}
\affiliation{Cahill Center for Astronomy and Astrophysics, California Institute of Technology, 1216 E California Blvd, Pasadena, CA 91125, USA}

\author[0000-0003-1252-4891]{Kristin K.\ Madsen}
\affiliation{CRESST and X-ray Astrophysics Laboratory, NASA Goddard Space Flight Center, Greenbelt, MD 20771 USA}

\author[0000-0003-4216-7936]{Guglielmo Mastroserio}
\affiliation{Cahill Center for Astronomy and Astrophysics, California Institute of Technology, 1216 E California Blvd, Pasadena, CA 91125, USA}

\author[0000-0001-5506-9855]{John A.\ Tomsick}
\affiliation{Space Sciences Laboratory, University of California, 7 Gauss Way, Berkeley, CA 94720-7450, USA}

\author[0000-0001-9110-2245]{Daniel Wik}
\affiliation{Department of Physics and Astronomy, University of Utah, 115 1400 E., Salt Lake City, UT 84112 USA}

\submitjournal{The Astrophysical Journal}
\accepted{17 Jan.\ 2022}
\received{12 Nov.\ 2021}
\revised{6 Jan.\ 2022}

\begin{abstract}
StrayCats, the catalog of NuSTAR stray light observations, contains data from bright X-ray sources that fall within crowded source regions. These observations offer unique additional data with which to monitor sources like X-ray binaries that show variable timing behavior. In this work, we present a timing analysis of stray light data of the high mass X-ray binary SMC X-1, the first scientific analysis of a single source from the StrayCats project. We describe the process of screening stray light data for scientific analysis, verify the orbital ephemeris, and create both time and energy resolved pulse profiles. We find that the orbital ephemeris of SMC X-1 is unchanged and confirm a long-term spin up rate of $\dot{\nu}=(2.52\pm0.03)\times10^{-11}$ Hz s$^{-1}$. We also note that the shape of SMC X-1's pulse profile, while remaining double-peaked, varies significantly with time and only slightly with energy.
\end{abstract}

\section{Introduction} \label{sec:int}
The Nuclear Spectroscopic Telescope ARray (NuSTAR, \citealt{harrison2013}) is the first satellite in orbit that can focus high energy X-rays between 3--79 keV. The telescope is constructed so that the optics are separated from the detectors by a 10 m open geometry mast. Because the mast is not enclosed, bright sources within 1--4\degree of the focused target can cause stray light, or ``aperture flux" to fall on the detector (see \citealt{madsen2017} for a complete description). \cite{grefenstette2021} presented \textbf{StrayCats}\footnote{https://nustarstraycats.github.io/}, the catalog of stray light observations and suggested that these observations could be useful in a variety of ways, including performing spectroscopy above 79 keV, reducing telemetry loads for bright sources, and obtaining extra coverage of sources. 

Stray light data offers a unique opportunity to increase the observation time of bright X-ray sources, such as accreting X-ray binaries. These sources often go through accretion state changes or exhibit fluctuations in their timing properties that make them interesting to monitor regularly. Stray light observations can offer serendipitous observations of these sources, thereby increasing the baseline of observations for these variable sources.

The high mass X-ray binary SMC X-1 is an interesting source with which to perform a scientific analysis of stray light data. This persistently bright source ($L_{\text{X}} \sim 3\times10^{38} \text{erg} \text{ s}^{-1}$) consists of a 1.21 \msun\ neutron star orbiting an 18 \msun\ B0 supergiant star (\citealt{liller1973,falanga2015}). The system is an eclipsing binary with a period of $\sim$ 3.9 days and a binary inclination of $\sim$ 70\degree\ (\citealt{schreier1972,vandermeer2007}). SMC X-1 is a frequent stray light source due to its X-ray luminosity and its proximity to other X-ray targets in the Small Magellanic Cloud. 

SMC X-1 exhibits X-ray pulsations from the rotation of the neutron star with a period of $\sim$ 0.7 s (\citealt{lucke1976}). Long term studies of the pulse behavior in this source have indicated that the spin frequency is increasing due to accretion torque at a rate of $\sim 2.6\times10^{-11}$ Hz s$^{-1}$ (\citealt{inam2010,hu2019}). SMC X-1 also shows an irregular superorbital period of 45-60 days that has long been thought to be caused by warping of the inner accretion disk that obscures the pulsar as the disk precesses (\citealt{wojdowski1998,clarkson2003,brumback2020}). However, more recently there has been evidence that absorption alone is not responsible for the superorbital period and some changes in intrinsic flux may be occurring as well (\citealt{pradhan2020}). The most typical superorbital period is about 55 days in length and has a simple sinusoidal shape that defines the source's high and low states. The duration of the superorbital cycle is quasi-periodic and excursions that bring the period down to 40 days occur roughly every 6 years (\citealt{wojdowski1998,clarkson2003,hu2019}). These excursions are potentially caused by an instability in the warping of the accretion disk, which changes its geometry as it moves between stable modes (\citealt{ogilvie2001,hu2019,dage2019}).These short- and long-term timing behaviors make SMC X-1 particularly interesting to examine through additional stray light observations.

In this work, we present the first timing analysis of NuSTAR stray light data through 8 observations of SMC X-1. We describe the observations, data reduction, and our screening criteria in Section \ref{sec:obs}. In Section \ref{sec:an} we first use stray light observations that show binary eclipses to confirm the accuracy of the SMC X-1 orbital ephmeris, and then examine the spin behavior and pulse profile shape within our observations. In Section \ref{sec:results} we discuss our results, including the long term spin behavior of SMC X-1. 

\section{Observations} \label{sec:obs}
\textbf{StrayCats} contains eight observations of SMC X-1 in stray light ranging from October 2015 to October 2018. We present a brief summary of these observations in Table \ref{tab:obs}, including both the NuSTAR ObsID and the Stray Light ID from \textbf{StrayCats}. We cleaned and analyzed the data using HEASoft v6.29c, NuSTARDAS v2.1.1, the NuSTAR CALDB v20210921, and the {\fontfamily{qcr}\selectfont nustar-gen-utils}\footnote{https://github.com/NuSTAR/nustar-gen-utils} software package. For all observations, we re-ran the NuSTAR pipeline {\fontfamily{qcr}\selectfont nupipeline} and extracted a detector image to determine the quality of the stray light.  We screened these observations for scientific suitability using the following criteria: the stray light pattern covered more than 1 cm$^{2}$ of the NuSTAR detector and is uncontaminated by overlapping stray light patterns from other nearby X-ray sources. 

\begin{figure*}
\centering
\includegraphics[scale=0.7]{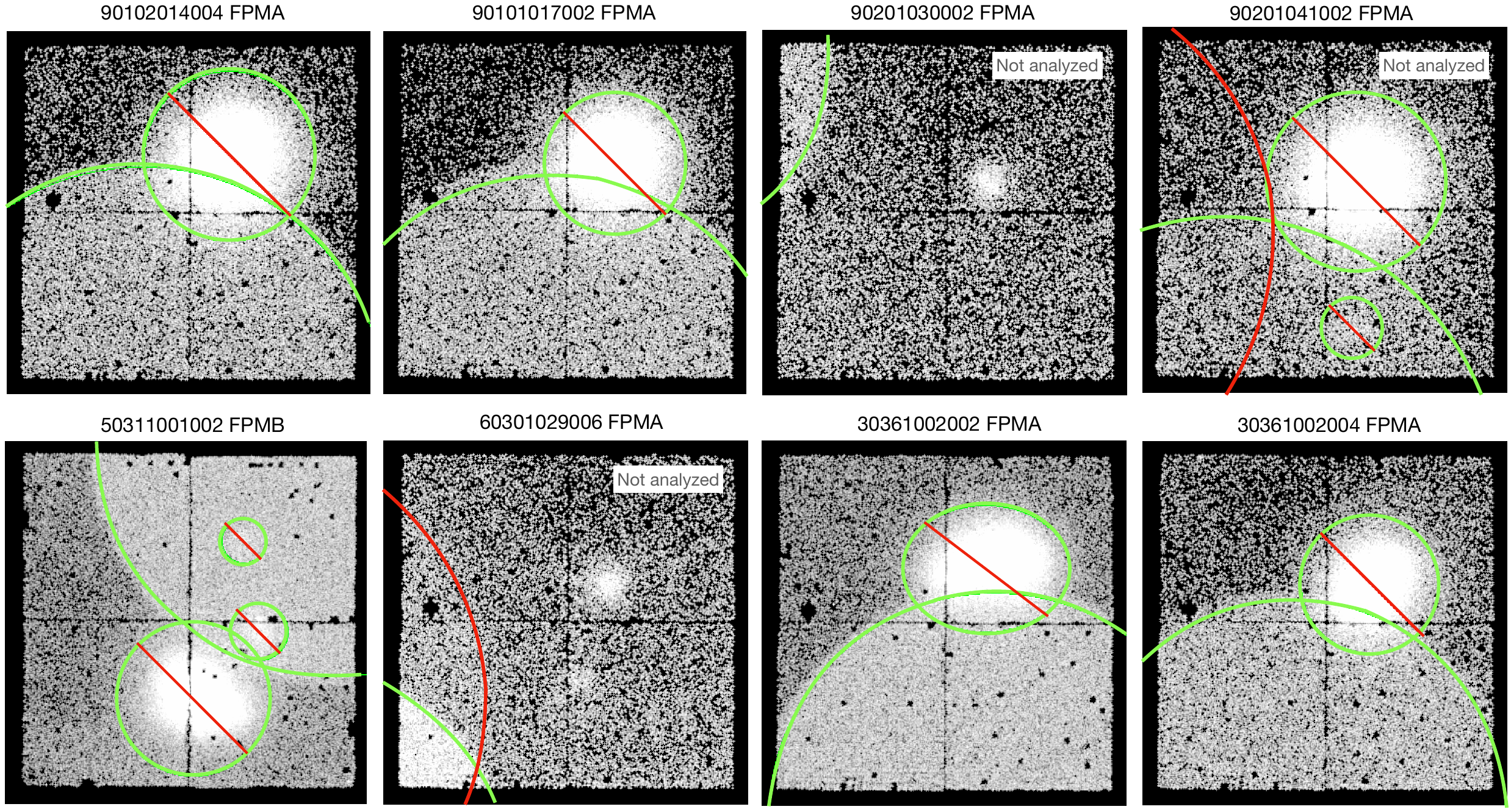}
\caption{NuSTAR 3--30 keV images in the Detector 1 reference showing the stray light pattern of SMC X-1 in all eight observations within \textbf{StrayCats}. Images have been smoothed with a Gaussian filter. Large green circles indicate the source regions that encompass the SMC X-1 stray light pattern in each observation. In observations where point sources, generally the focused target, overlap the stray light pattern, we remove this emission via exclusion regions (green circle or ellipse with red line). Stray light patterns from sources other than SMC X-1 are shown in red circles. We determined that three of the eight observations were unsuitable for scientific analysis either because the stray light pattern was too small or because of contamination from other stray light sources. }
\label{fig:regions}
\end{figure*}

Following this screening, we found that three of these observations are unsuitable for scientific analysis. Observation 90201030002 shows an extremely small stray light pattern from SMC X-1 that takes up less than 1 cm$^{2}$ on the NuSTAR detector, resulting in too few counts for a scientific analysis. Observationss 90201041002 and 60301029006 both have more sizable stray light patterns, but those patterns are contaminated by overlapping stray light emission from other sources. We show the NuSTAR detector images for these observations and highlight their complications in Figure \ref{fig:regions}.

For the five remaining observations, we created circular source regions in DS9 that encompassed the stray light pattern on the NuSTAR detector (see Figure \ref{fig:regions}). When necessary, we also created circular or elliptical exclusion regions centered over the focused target to isolate emission from SMC X-1. We determined the sizes of these exclusion regions visually to encompass the area where the PSF surface brightness of the source reached the background level. We performed the barycentric correction on the source-selected event files using the tool {\fontfamily{qcr}\selectfont barycorr} with the DE-200 ephemeris and provided the right ascension and declination of SMC X-1, not the focused target.

\section{Analysis} \label{sec:an}

\begin{figure*}
\centering
\includegraphics[scale=1.2]{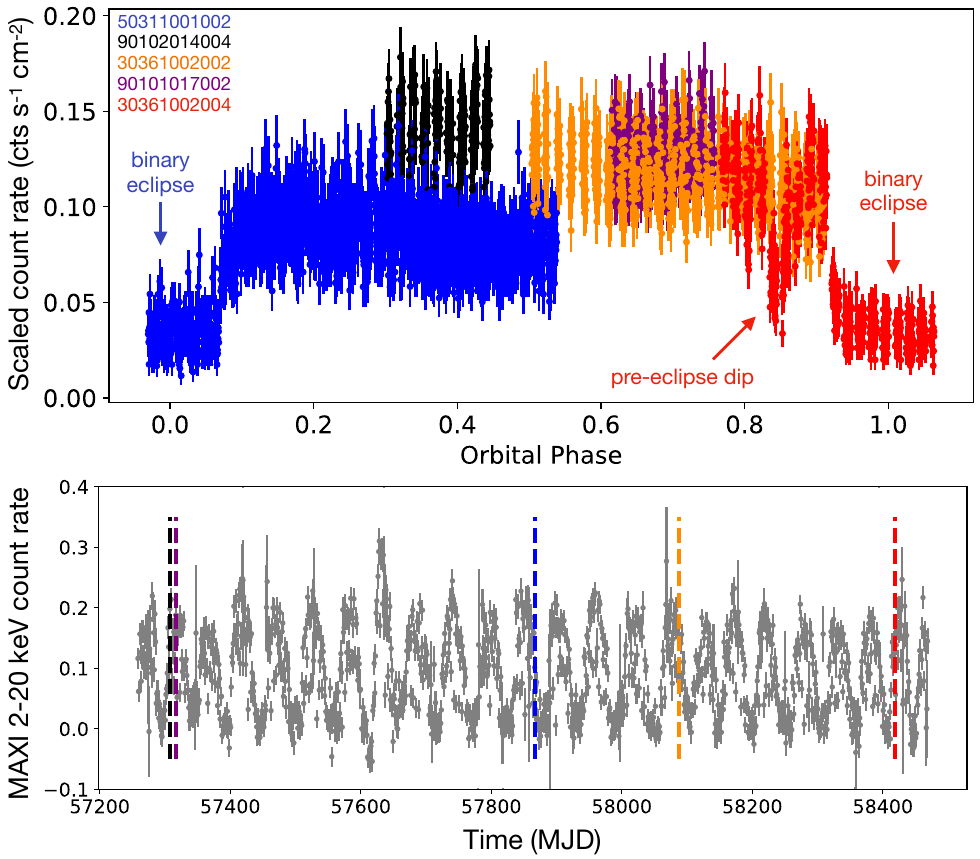}
\caption{Top: NuSTAR 3-30 keV light curves binned by 100 s for Observations 90102014004 (black), 90101017002 (purple), 50311001002 (blue), 30361002002 (orange), and 30361002004 (red) plotted as a function of orbital phase. The count rates have been scaled by the areas of the respective stray light source regions to highlight source variability. We filtered the light curves to only plot bins with exposure fractions greater than 0.8. Observation 50311001002 and 30361002004 show parts of SMC X-1's binary eclipse. Additionally, Observation 30361002004 shows a pre-eclipse dip. Bottom: The MAXI 2--20 keV long term light curve (grey) of SMC X-1 during the three year interval covered by our observations. The times of stray light observations are indicated with colored dashed lines, where the colors correspond to those used in the top panel.}
\label{fig:lcs}
\end{figure*}

We extracted light curves, source filtered event files, and source spectra using {\fontfamily{qcr}\selectfont nustar-gen-utils}. We show the NuSTAR 3-30 keV light curves, binned by 100s, in Figure \ref{fig:lcs} as a function of orbital phase. The light curves for Observations 50311001002 and 30361002004 show parts of SMC X-1's binary eclipse. Additionally, Observation 30361002004 also captured a pre-eclipse dip. Such dips have been observed previously in SMC X-1 (\citealt{hu2013}) and in a similar accreting pulsar, Her X-1 (\citealt{giacconi1973}). In both sources, such dips are likely caused by obscuration of the neutron star by the impact region where the accretion disk meets the accretion stream (\citealt{hu2013}).

Before beginning pulsation searches, we first examined where each of the five observations occurred within SMC X-1's superorbital cycle. SMC X-1's X-ray pulsations are often undetected during the superorbital low state due to increased obscuration of the neutron star by the accretion disk (e.g.\ \citealt{pike2019,brumback2020}). By examining long term MAXI (\citealt{maxi}) light curves around the time of our observations, we determined that all five of our observations occurred within SMC X-1's superorbital high state, meaning that we could reasonably expect to detect pulsations if the stray light data provided high enough signal to noise.

\begin{splitdeluxetable*} {cccccccBccc}
\label{tab:obs} 
\tablecolumns{10}
\tablecaption{Description of SMC X-1 Stray Light Observations} 
\tablewidth{0pt}
\tablehead{
\colhead{ObsID} & \colhead{Stray Light ID} & \colhead{Date} & \colhead{Start Time (MJD)} & \colhead{Exposure (ks)} & \colhead{Module} & \colhead{Stray light area (cm$^{2}$)} & \colhead{ObsID} &\colhead{Pulse Frequency (Hz)} & \colhead{$F_{\text{3--30 keV}}$ (erg cm$^{-2}$ s$^{-1})$} }
\startdata
90102014004 & 10 & 12 Oct.\ 2015 & 57307.91 & 23.1 & FPMA & 7.68 & 90102014004 & 1.428669(2) & ($1.91 \pm 0.05)\times 10^{-9}$ \\
90101017002 & 9 & 21 Oct.\ 2015 & 57316.91 & 26.7 & FPMA & 7.73 & 90101017002 & 1.4286958(9) & ($1.98 \pm 0.04)\times 10^{-9}$ \\
90201030002 & 2 & 17 Jul.\ 2016 & 57586.72 & 54.8 & FPMA/B & 0.69 & 90201030002 & Not Analyzed\tablenotemark{a} & \nodata \\
90201041002 & 3 & 12 Nov.\ 2016 & 57704.76 & 38.6 & FPMA & 4.31 & 90201041002 & Not Analyzed\tablenotemark{b} & \nodata \\
50311001002 & 11 & 24 Apr.\ 2017 & 57867.06 & 153.2 & FPMB & 5.53 & 50311001002 & 1.429871(2) & ($1.21 \pm 0.02)\times 10^{-9}$ \\
60301029006 & 8 & 17 Nov.\ 2017 & 58074.69 & 74.4 & FPMA & 0.66 & 60301029006 & Not Analyzed\tablenotemark{a,b} & \nodata \\
30361002002 & 4 & 30 Nov.\ 2017 & 58087.04 & 70.7 & FPMA & 7.08 & 30361002002 & 1.430399(3) & ($1.49 \pm 0.02)\times 10^{-9}$ \\
30361002004 & 5 & 27 Oct.\ 2018 & 58418.87 & 58.3 & FPMA & 6.97 & 30361002004 & 1.431085(2) & ($1.65 \pm 0.03)\times 10^{-9}$ \\
\enddata
\tablenotetext{a}{Stray light pattern covers $<1 \text{ cm}^{2}$ on detector}
\tablenotetext{b}{SMC X-1's stray light pattern is contaminated by the stray light patterns of other sources}
\end{splitdeluxetable*}

\subsection{Verifying the orbital ephemeris}
As shown in Figure \ref{fig:lcs}, two of our five observations show binary eclipses in their NuSTAR light curves. To our knowledge, the most recent orbital ephemeris for SMC X-1 was determined by \cite{falanga2015} (hereafter F15) with an update to the rate of orbital decay provided by \cite{hu2019}. The rate of orbital decay provided by \cite{hu2019} differed from the F15 value by 4.7\%. Because the two eclipses observed in stray light extended the baseline of observed eclipses from those presented in F15, we tested this ephemeris to see if the orbital parameters had changed. We performed a preliminary test of both the F15 and \cite{hu2019} ephemerides and found that, for some observations, the F15 ephemeris produced stronger detected pulsations. For this reason, we proceeded with verifying the F15 ephemeris.

We measured the time of the end of the eclipse in Observation 50311001002 and the start time of the eclipse in Observation 30361002004 as the time when 99\% of the NuSTAR flux was obscured. We calculated the time of mid-eclipse using these values and the assumption that the eclipse duration lasts 0.127 of the orbital phase (F15) and found that the times of mid-eclipse were $57867.182 \pm 0.032$ MJD and $58419.775 \pm 0.025$ MJD. It is worth noting that the eclipses used by F15 were observed with INTEGRAL (\citealt{integralisgri}), which has a harder energy range compared to NuSTAR, and thus might have a slightly shorter eclipse duration. We added these two data points to those presented in Appendix A of F15 for SMC X-1.  

\begin{figure*}
\centering
\includegraphics[scale=0.9]{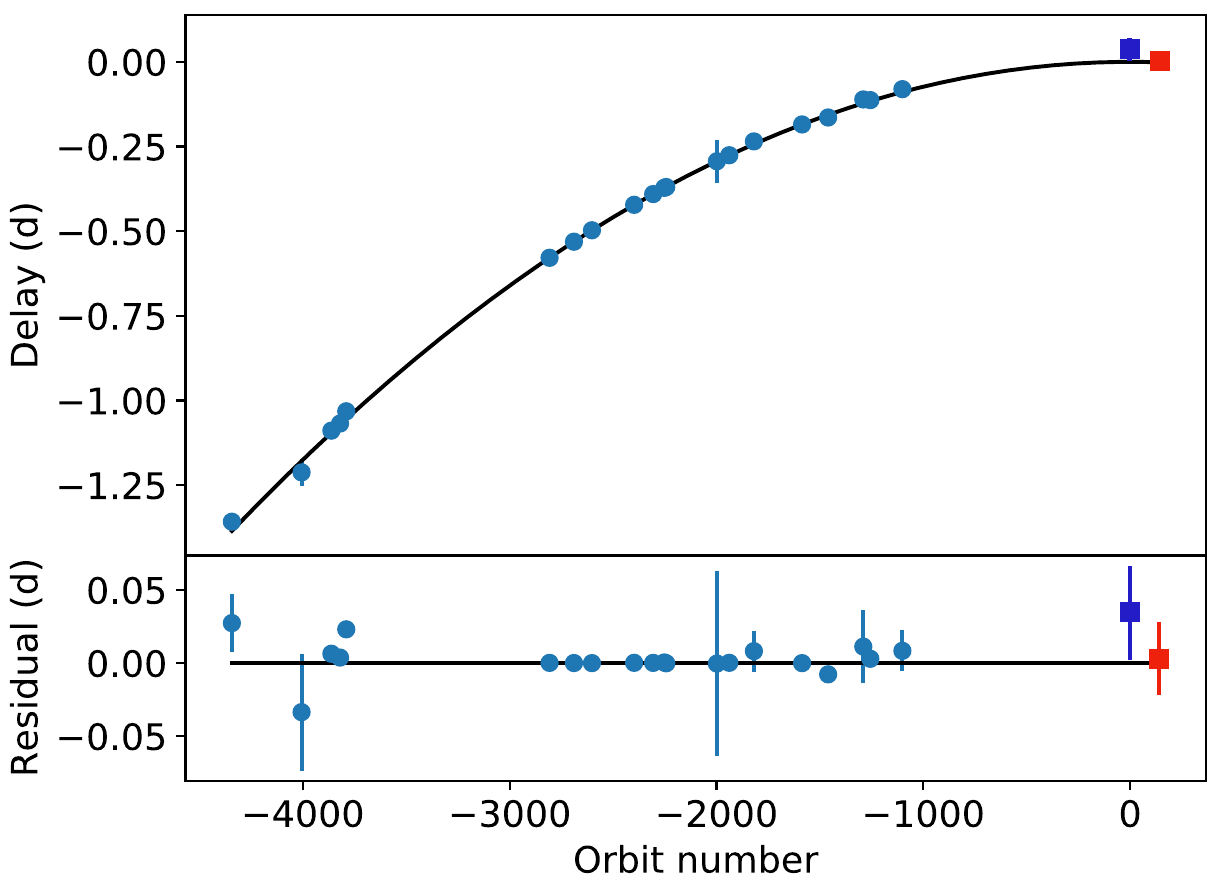}
\caption{Top: the time delay between the SMC X-1 observed times of mid-eclipse and the times predicted by the linear terms of the quadratic orbital change equation. Light blue circles are data points from \cite{falanga2015} and the squares are the two eclipses observed in stray light (Observation 50311001002 in blue and Observation 30361002004 in red). The colors of the squares match the colors of the observation light curves in Figure \ref{fig:lcs}. The black line is the best fit quadratic equation, which is consistent with the \cite{falanga2015} ephemeris. Bottom: Residuals to the above fit.}
\label{fig:orb}
\end{figure*}

Following the work of F15, we use the quadratic orbital change function to predict the time ($T_{n}$) of an eclipse an integer number $n$ times after the reference time $T_{0}$:
$$T_{n}=T_{0} + Pn + \frac{1}{2}P\dot{P}n^{2}$$
where $P$ and $\dot{P}$ are the orbital period and the orbital period first derivative, respectively. In order to isolate and examine the quadratic term and the change in $\dot{P}$, we plotted the time delay between the observed eclipses and the linear terms of the equation above (Figure \ref{fig:orb}). We selected $T_{0}$ as the mid-eclipse time of our first observed stray light eclipse (57867.182 MJD) and fit the time delay data with a quadratic equation, allowing us to solve for the orbital parameters. 

We find that the orbital parameters calculated from our fit shown in Figure \ref{fig:orb} are consistent with those of F15 forward-modeled to the appropriate time. With this consistency, we confirm that the F15 ephemeris for SMC X-1 is still an accurate description of the source's orbital behavior. 

\subsection{Selecting energy ranges}
In order to select the most appropriate energy ranges over which to perform our timing analysis, we examined the 3--100 keV spectra for our five observations. Because stray light does not pass through NuSTAR's optics, we can examine energies above the calibrated range for focused sources with the accuracy determined by the background and the detector calibration. The reduced signal to noise in the stray light observations meant that features such as the Fe K$\alpha$ line, which is commonly observed in this source with NuSTAR (e.g.\ \citealt{brumback2020}), were not detected.

For each observation, a circular background region was selected on whichever NuSTAR detector allowed us to find area away from the focused target and stray light signals. In all observations, we found that the stray light spectrum of SMC X-1 became background dominated around 30 keV. This value represents a weaker overall source spectrum with respect to the background compared to focused NuSTAR spectra of this source, where the background begins to dominate $\sim$50 keV (e.g.\ \citealt{brumback2020}). As with the lack of an obvious Fe K$\alpha$ line, this is likely due to the reduced signal in stray light.

\begin{figure}
\centering
\includegraphics[scale=0.8]{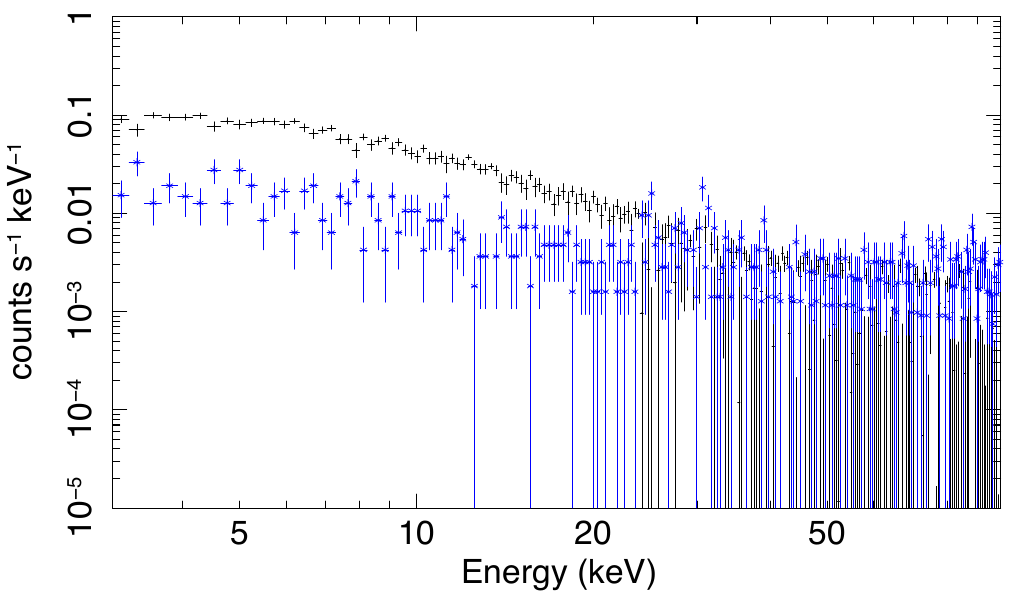}
\caption{The NuSTAR 3--100 keV stray light spectrum (black) for Observation 90101017002, as a representative example. The source spectrum becomes dominated by the background (blue) around 30 keV, thus setting an upper energy limit on our timing analysis. The 3--30 keV flux for this and the other analyzed observations can be found in Table \ref{tab:obs}. }
\label{fig:spectrum}
\end{figure}

To illustrate the effect of the background in stray light, we show a representative spectrum from Observation 90101017002 in Figure \ref{fig:spectrum}. Examining the point where the spectrum became background dominated allowed us to determine that we would not search for pulsations above 30 keV. 

For each of our five observations, we applied a coarse spectral model of an absorbed power law ({\fontfamily{qcr}\selectfont tbnew*cutoffpl}) with which to calculate the 3--30 keV flux of each observation. We present these flux values in Table \ref{tab:obs}. We defer a more detailed spectral analysis to future work.

\subsection{Pulsar spin frequency analysis}
Following our confirmation of the F15 orbital ephemeris for SMC X-1, we used this solution to correct our observations for its orbital motion.  

\begin{figure}
\centering
\includegraphics[scale=1.4]{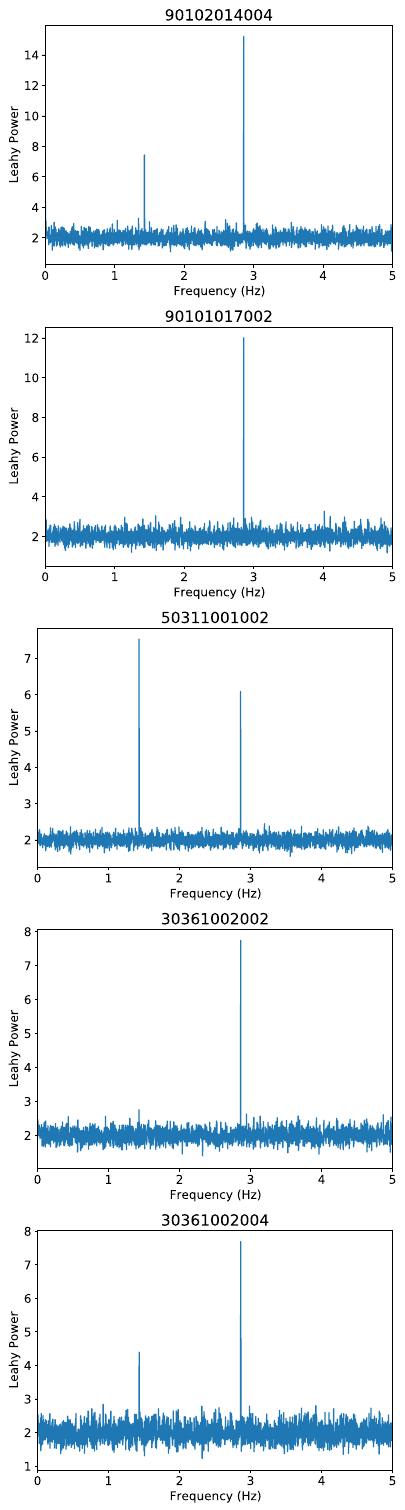}
\caption{Power density spectra for NuSTAR in the 3-30 keV energy range. In each observation, pulsations at the $\sim$1.4 Hz spin frequency and/or its $\sim$2.8 Hz harmonic are seen. The harmonic frequency features strongly in SMC X-1's power density spectra because the pulse profile is double peaked.}
\label{fig:pds}
\end{figure}

We used {\fontfamily{qcr}\selectfont nustar-gen-utils} to extract event files in the 3--30 keV energy band for each observation for our pulsation search. Using the timing softwares HENDRICS (\citealt{hendrics}) and Stingray (\citealt{stingray}), we created power density spectra (see Figure \ref{fig:pds}) that indicated pulsations were detected around 1.4 Hz (or the 2.8 Hz harmonic, which is frequently seen in SMC X-1 because of its double-peaked pulse profile) in each observation. We used the HENDRICS epoch folding search to find the best fit spin frequency and spin frequency first derivative. In all observations, the distribution on the spin frequency first derivative was consistent with zero. We list our best fit spin frequencies in Table \ref{tab:obs}. We determined the uncertainty on the spin frequency by using Xselect to extract short time intervals from the start and end of each observation. We created pulse profiles from these start and end time intervals and measured the change in pulse phase between the two. We then calculated our uncertainty in spin frequency as $\delta\nu = \delta\text{phase}/\delta\text{time}$.

\begin{figure}
\centering
\includegraphics[scale=1.4]{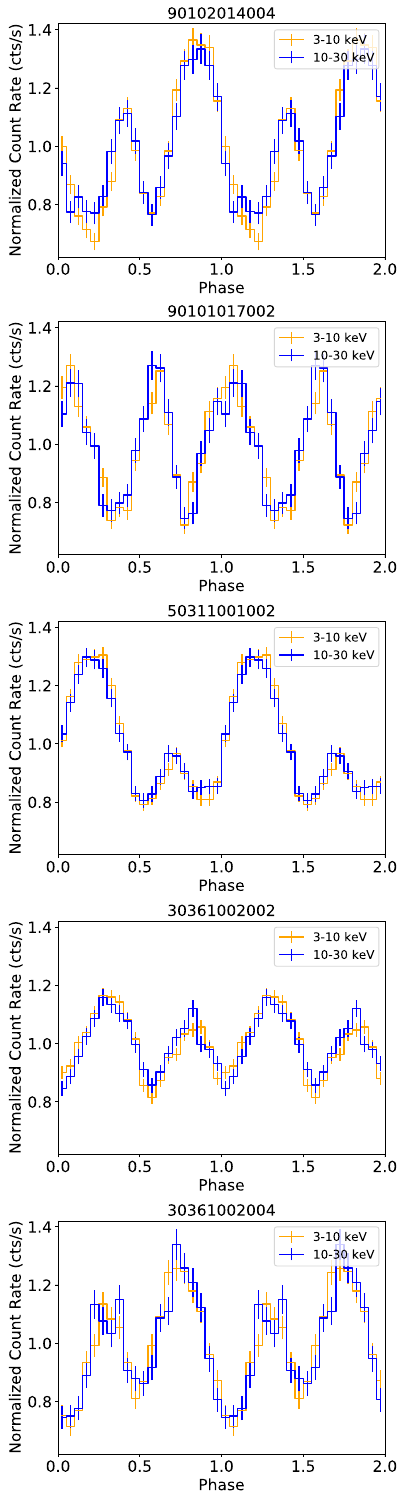}
\caption{NuSTAR 3-10 keV (orange) and 10-30 keV (blue) pulse profiles for the five observations. These profiles are not phase linked. The y-axis scale is the same in each plot to demonstrate changes in pulsation strength.}
\label{fig:pp}
\end{figure}

The entire observation was used to make the pulse profile except in the case of Observation 50311001002 and Observation 30361002004, in which the data during eclipse was excluded from the pulse profiles. We used Xselect to filter the event files into two energy ranges, 3--10 keV and 10--30 keV, in order to examine energy dependent changes. We used the {\fontfamily{qcr}\selectfont fold\_events} tool from Stingray (\citealt{stingray}) to create the pulse profiles, which are shown in Figure \ref{fig:pp}. 

\begin{figure*}
\centering
\includegraphics[scale=0.95]{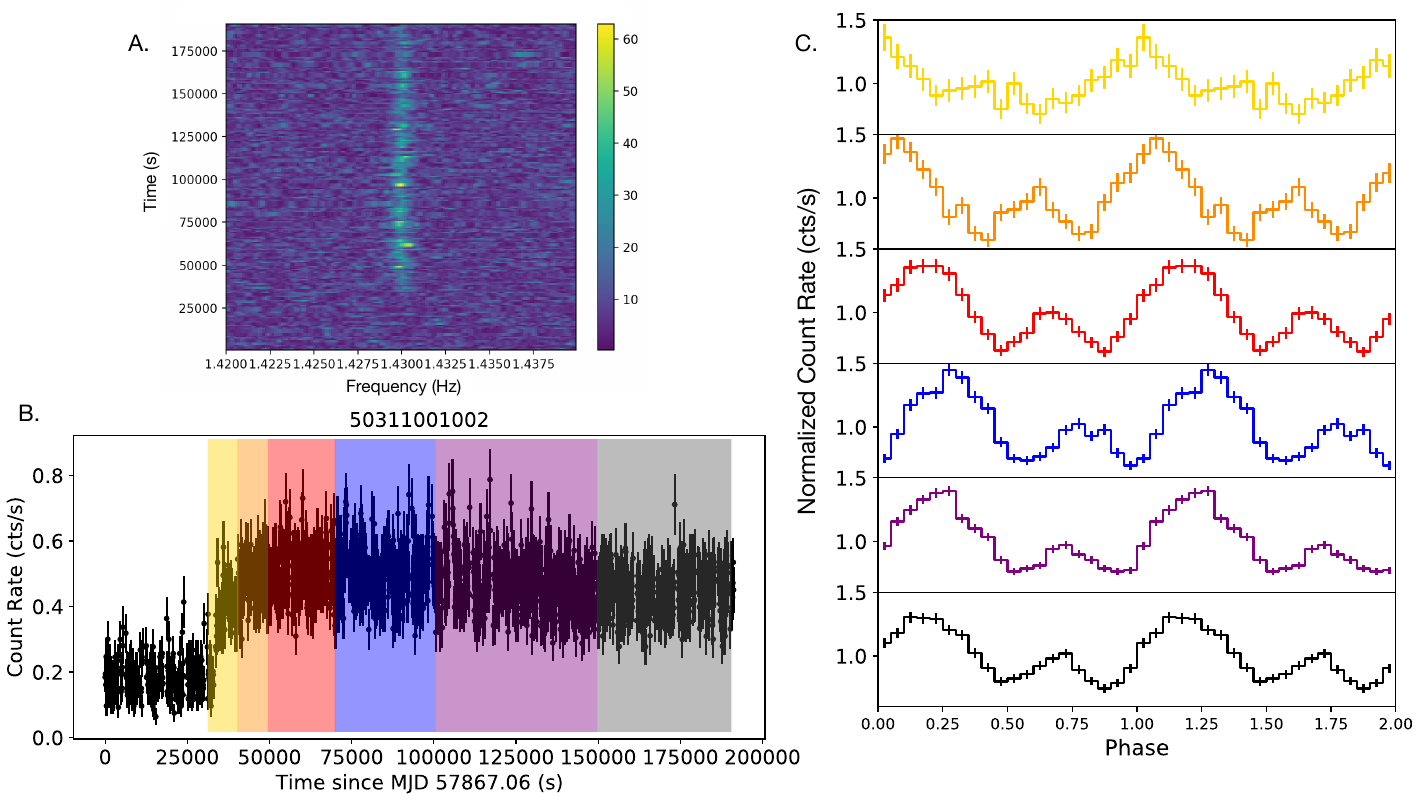}
\caption{Panel A: A dynamical z-search of Observation 50311001002 with HENDRICS, where the color indicates the strength of the pulsations with Z$^{2}_{n}$ statistics (\citealt{buccheri1983,bachetti2021}). The egress from eclipse shows some variations in the strongest pulse frequency, indicating change in the pulse shape. Panel B: The light curve of Observation 50311001002 where the colored panels indicate the time intervals from which the pulse profiles in Panel C were extracted. Panel C: Pulse profiles in the NuSTAR 3--30 keV band for each time interval.}
\label{fig:502}
\end{figure*}

In the case of Observation 50311001002, our timing analysis indicated small changes in the pulse profile shape as SMC X-1 emerged from binary eclipse. We used Xselect to filter this observation into several shorter time intervals in order to examine the time dependence of the pulse profiles in this observation (Figure \ref{fig:502}). We discuss these results in Section \ref{sec:results}.

\begin{figure*}
\centering
\includegraphics[scale=0.7]{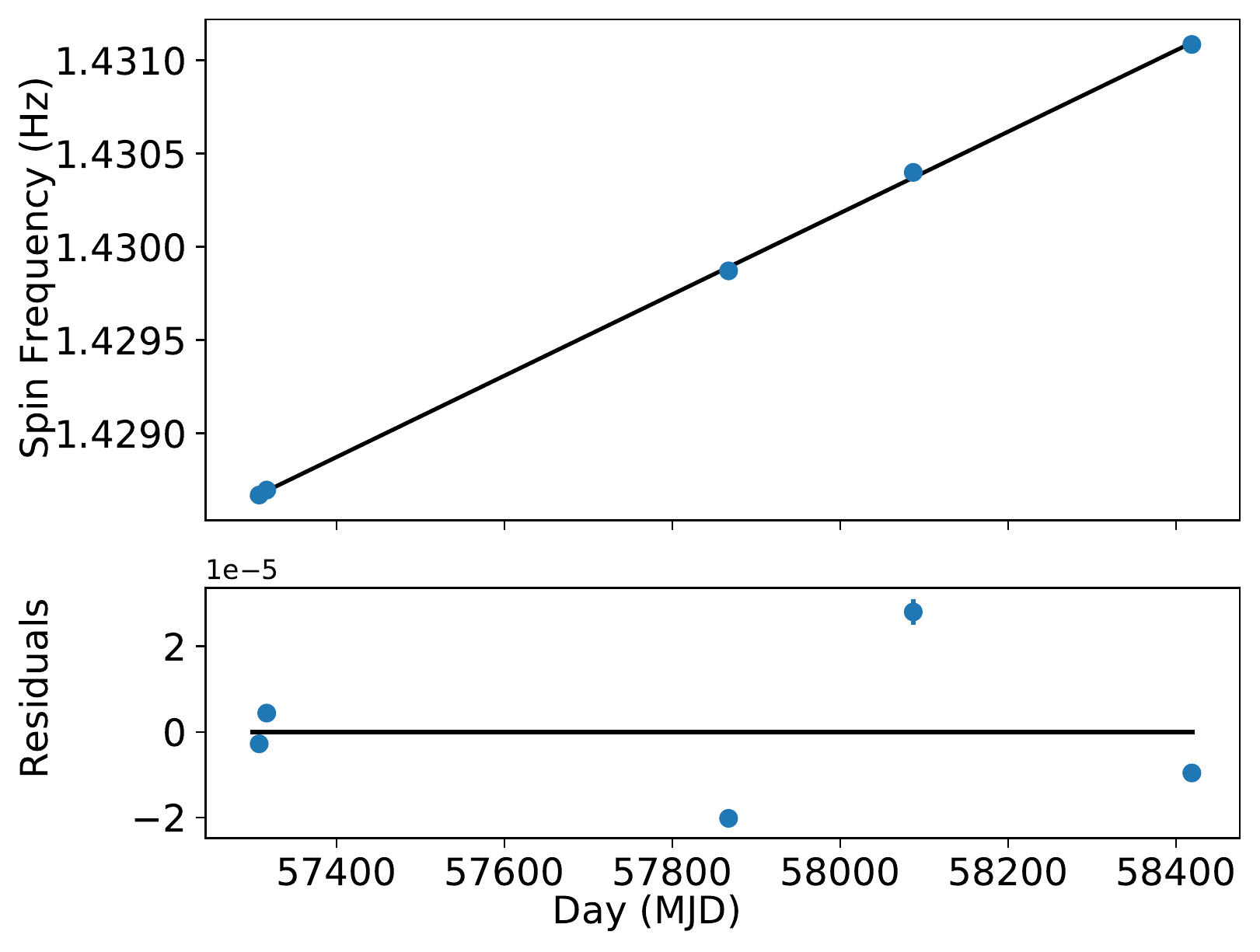}
\caption{Top: The best fit linear regression (black) to our spin frequency data from Table \ref{tab:obs} (blue points). The fit indicates a spin up of $\dot{\nu}=(2.52\pm0.03)\times10^{-11}$ Hz s$^{-1}$ with a p-value of $2.1 \times10^{-6}$. The error bars are with the point markers. Bottom: Residuals to the above fit. }
\label{fig:spinup}
\end{figure*}

The values in Table \ref{tab:obs} indicate that SMC X-1 is spinning up over the course of these observations. We preformed a linear regression on the spin frequencies presented in Table \ref{tab:obs} and found they indicate a spin up rate of $\dot{\nu}=(2.52\pm0.03)\times10^{-11}$ Hz s$^{-1}$. This value was determined using a least-squares linear regression, which produced a p-value of $2.1\times10^{-6}$. We show this fit and the resulting residuals in Figure \ref{fig:spinup} and discuss the implication in Section \ref{sec:results}.

\section{Results and Discussion} \label{sec:results}

The pulse profiles extracted from our stray light observations of SMC X-1 (see Figure \ref{fig:pp}) all show the double-peaked profile that is characteristic of this source. We do see some variations between observations, mainly in the relative strength of the weaker secondary pulse. Many previous works examining SMC X-1's pulse profiles have noted similar changes in the pulse shape with time (e.g.\ \citealt{neilsen2004,hickoxvrtilek2005,raichur2010}). In Figure \ref{fig:pp} we show pulse profiles from the 3--10 keV and 10--30 keV energy bands for each observation, although we only see slight differences between the two energy bands. This is consistent with previous studies of energy-resolved pulse profiles in SMC X-1; the double-peaked structure is typically present across the X-ray band, but single peaked pulse profiles have been seen in the soft X-rays (below 1 keV) where reprocessed emission from the accretion disk dominates (\citealt{hickoxvrtilek2005,brumback2020}).

Observation 50311001002 stands out from the others by showing a pulse profile with a significantly weaker secondary pulse (Figure \ref{fig:pp}). A further investigation revealed variations in pulse profile shape as the pulsar emerges from behind its companion star. We first see this behavior as irregularities in the dynamical z-search of the observation between $3-6\times10^5$ s in the observation, as the source brightens. By extracting pulse profiles in short time intervals over the course of the observation, we see changes in the pulse profile shape from single peaked to double peaked as well as an overall phase shift (Figure \ref{fig:502}, Panel C). These changes could be driven by the changing view of the pulsar as it emerges from eclipse, although further modeling is needed to confirm this scenario. 

Spin up has been well documented in SMC X-1 since the source's discovery and is thought to be driven by accretion torques (e.g.\ \citealt{hu2019}). The long term spin behavior is generally linear, although the rate of spin up appears to vary slightly with time (\citealt{inam2010,hu2019,pradhan2020}). The cause of these variations is not well understood because there are no clear indicators of a change in mass accretion rate (e.g.\ \citealt{inam2010,hu2019}). \cite{dage2019} saw a tentative correlation between spin up rate and the superorbital period length, possibly due to changing accretion flow that affected both the disk's structure and the pulsar's accretion rate. This conclusion was largely driven by an observed change in SMC X-1's spin up behavior around MJD 50000, which was followed by an epoch of superorbital excursion (\citealt{dage2019,pradhan2020}).

Our observed spin up value of $\dot{\nu}=(2.52\pm0.03)\times10^{-11}$ Hz s$^{-1}$ is consistent with the spin up estimations from \cite{hu2019} and \cite{pradhan2020}. This agreement indicates that timing analysis of other stray light sources could be used to extend the baseline of pulsation monitoring in similar sources.

\section{Conclusions} \label{sec:con}

In this work, we have examined all stray light observations for the high mass X-ray binary SMC X-1 in the NuSTAR stray light catalog \textbf{StrayCats}. We determined that three of the eight available observations were unsuitable for scientific analysis due to either a small stray light pattern on the NuSTAR detector or contamination from other stray light sources.

For our remaining five observations, we used the {\fontfamily{qcr}\selectfont nustar-gen-utils} software to extract data products and performed a timing analysis with these data. Because two observations showed binary eclipses in their light curves, we added these eclipse times to those presented in F15 and confirmed that the orbital ephemeris for SMC X-1 remains unchanged. We performed a pulsation search and found the best fit spin frequency for each observation, noting a spin-up of $\dot{\nu}=(2.52\pm0.03)\times10^{-11}$ Hz s$^{-1}$ across our observations, which is consistent with calculations of the spin-up in this source from focused observations (\citealt{hu2019,pradhan2020}). 

We created pulse profiles in the 3--10 and 10--30 keV energy bands for our five observations (see Figure \ref{fig:pp}), and explored the time dependence of the pulse profiles in the observation that showed the most unique pulse shape (see Figure \ref{fig:502}). We noted that there are significant variations in pulse shape with time, likely caused by the neutron star emerging from binary eclipse. 

This is the first scientific analysis of a single source from the \textbf{StrayCats} project. We have demonstrated the ability of NuSTAR stray light observations to obtain good timing solutions and pulse profiles from these unique observations. In particular, stray light observations offer the opportunity to extend baseline monitoring in sources with variable timing properties.

\begin{acknowledgements}
The authors thank the anonymous referee for their time in reading the manuscript and providing helpful comments. MCB would like to thank J.\ Thorstensen for communication about calculating orbital ephemerides and acknowledges support from NASA funding award number 80NSSC21K0263. RK acknowledges support from the Russian Science Foundation grant 19-12-00369. RL acknowledges the support of NASA through the Hubble Fellowship Program grant HST-HF2-51440.001.
\end{acknowledgements}

\software{HEAsoft (v6.29c; HEASARC 2014), NuSTARDAS, Stingray (\citealt{stingray}), HENDRICS (\citealt{hendrics}), MaLTPyNT (\citealt{maltpynt}), {\fontfamily{qcr}\selectfont nustar-gen-utils}}

\pagebreak

\bibliography{my_bib.bbl}{}
\bibliographystyle{aasjournal}

\end{document}